# Heat transfer in γ-phase of oxygen

V. A. Konstantinov, V. G. Manzhelii, V. P. Revyakin, and V. V. Sagan

Institute for Low Temperature Physics & Engineering of NASU, 61103, 47 Lenin Ave., Kharkov, Ukraine.

The isochoric thermal conductivity of γ-$O_2$ has been studied on samples of varying density in the temperature interval from 44 K to the onset of melting. The thermal conductivity of nearly free sample decreases at rising temperature along the isochores. It is shown that the absolute value of thermal conductivity in the γ- phase of $O_2$ is close to its lower limit and most of the heat is transported by «diffusive» modes. The growth of thermal conductivity in γ-$O_2$ is attributed to the decay of the phonon scattering at the rotational excitations of the molecules and at the short-range magnetic order fluctuations at rising temperature.

**Introduction**

Solid oxygen belongs to a small group of molecular crystals consisting of linear molecules. In contrast to the $N_2$-class crystals having the orientationally ordered *Pa3*- type structure, solid oxygen, much like halogens, has a collinear orientational packing because the valence rather than quadrupole, forces predominate in the noncentral interaction. Besides, in the ground state the $O_2$ molecule has the electron spin S=1 which determines the magnetic properties of oxygen. Another specific feature of solid $O_2$ is the fact that the energy of the magnetic interaction energy makes up a considerable portion of the total binding energy [1-2]. This unique combination of molecular parameters has stimulated much interest in the physical properties of $O_2$, in particular its thermal conductivity.

The thermal conductivity of solid $O_2$ was investigated under saturated vapor pressure in the α-, β- and γ- phases in a temperature interval of 1-52 K [3-4]. The low-temperature α- $O_2$ phase is ordered orientationally and magnetically. It has a monoclinic face-centered structure with the spatial symmetry *C2/m* in which the axes of the molecules are collinear and perpendicular to the layers of the close-packing <0,0,1>. On heating to 23.9 K, the structure changes into the rhombohedral magnetically-ordered β-phase of the symmetry *R3m*. This is the simplest orientational structure, similar to α-$O_2$. On a further heating, orientational disordering (cubic cell, *Pm3m* symmetry with Z=8) occurs at T= 43.8 K[1-2]. Under atmospheric pressure oxygen melts at 54.4 K.

The thermal conductivity has a maximum in the α- phase at T≈6 K, drops sharply on a change to the β - phase, where it is practically constant, jumps again at the β→γ transition and increases in γ-$O_2$ [3-4]. The experimental results are interpreted as follows. In the magnetically ordered α- phase the heat is transferred by both phonons and magnons, and their contributions are close in magnitude: $\Lambda_{ph} \approx \Lambda_m$. On the α→β transition the thermal conductivity decreases sharply (~60%) because the magnon component disappears during magnetic disordering. The weak temperature dependence of the thermal conductivity in the β-phase is attributed to the anomalous temperature dependence of the sound velocity in β-$O_2$ which is practically constant for the longitudinal modes and increases for the transverse ones [1, 5]. The growth of the thermal conductivity in γ-$O_2$ is attributed to the decay of the phonon scattering at the rotational excitations of the molecules and at the short-range magnetic order fluctuations at rising temperature.

In the context of this quite full information about the features of the heat transfer in solid oxygen, some aspects are worthy of special attention. For a correct comparison with theory, the thermal conductivity should be measured at a constant density to exclude the effect of thermal expansion. This requirement is particularly important at rather high temperatures at which the thermal expansion coefficients are large. The growth of thermal conductivity was observed earlier in orientationally disordered phases of some molecular crystals [6-7]. True, under saturated vapor pressure such growth was observed, apart from oxygen, only in methane [8], where the molecular rotation becomes nearly free at the pre-melting temperatures. According to the calculation based on [3], γ-$O_2$ belongs to the molecular crystals which have the highest temperature coefficients of the growth of thermal conductivity f = $(\partial \ln \Lambda / \partial \ln T)_V$ (cf. f ≈ 3.5 for γ-$O_2$, ≈ 0.35 for $CH_4$ (I) and ≈ 0.5 for $CCl_4$ (I) [7]). This behavior is not quite clear in view of the fact that the molecular rotation in γ-$O_2$ is far from being free. Finally, above 35 K the phonon mean free path estimated from a simple gas-kinetic formula is close (≈ $10^{-7}$ cm [3]) to the lattice parameter, which makes the applicability of the phonon concept questionable. The above features have attracted our interest and stimulated this investigation of the isochoric thermal conductivity of γ-$O_2$.

**Experimental technique.**

Constant-volume investigations are feasible in molecular solids with comparatively low thermal pressure coefficient $(\partial P / \partial T)_V$. Samples of sufficient density can be grown in a high-pressure cell. Such samples can be cooled practically without changing their volumes, while the pressure in the cell is decreasing slowly. For samples of moderate densities, the pressure can drop to zero at a certain characteristic temperature $T_0$ and this violates the isochoric condition. On further cooling, the sample can separate from the cell walls. When the volume of the sample is constant, melting occurs in a certain temperature interval. Simultaneously, the onset of melting $T_m$ shifts towards higher temperatures as the density increases. This is clearly seen in the V-T phase diagram of solid $O_2$ plotted according to [2] (Fig. 1). $T_0$ (the onset of the single-phase region) shifts towards higher temperatures slower than $T_m$ does. As a result, the γ-region increases. The deviations from the constant volume caused by the thermal and elastic deformation of the measuring cell were usually no more than 0.3% and could be readily taken into account.

This investigation was made by a steady-state technique using a coaxial-geometry setup. The measuring beryllium bronze cell was 160 mm long with the inner diameter 17.6 mm. The diameter of the inner measuring cylinder was 10.2 mm. The highest permissible pressure was 800 MPa. The sensors of temperature were platinum resistance thermometers placed in special channels of the inner and outer cylinders to exclude a high pressure effect. The measuring cell had a temperature gradient of 1-2 K/cm during the growth of the samples. The pressure in the inflow capillary was varied within 50-200 MPa to produce samples of different densities. When the growth was completed, the capillary was blocked (frozen with liquid hydrogen) and the samples were annealed at pre-melting temperatures for one or two hours to remove density gradients. After measurement, the samples were evaporated into a thin-wall vessel and their masses were estimated by weighing. The molar volumes of the samples were found from the known volume of the cell and the sample mass. The total systematic measurement error dominant did not exceed 4% for the thermal conductivity and 0.2% for volume. The gas purity was no worse than 99.95%.

**Results and discussion**

The thermal conductivity was investigated on nearly free sample 4 grown under pressure ≈ 5 MPa ($V_m$=23.4 cm$^3$/mole). The isochoric thermal conductivity was studied on three samples of different densities in the temperature interval from 44 K to the corresponding melting temperatures. Sample 1 was grown under the ultimate pressure obtainable from the thermocompressor (about 180 MPa). As estimation shows, because of the large density jump during the β→γ transition, isochoric investigations of the thermal conductivity in β- phase are possible only if the growth pressure exceeds 300 MPa.

The experimental results are shown in Fig. 2. $T_0$ (the onset of the single-phase region) and $T_m$ (the onset of melting) are marked with arrows. These data, along with the molar volumes of the samples, are available in Table 1 and Fig. 1. The dashed line shows the thermal conductivity measured under saturated vapor pressure (see [3]). It is seen that our results for nearly free sample 4 agree in the absolute value with the data of [3] but they exhibit a different temperature dependence. Sample 4 was grown under the pressure ~5 MPa with a temperature gradient of 1.5 K over the measuring cell. The cooling rate was about 12 K/h. The sample was cooled to T~25 K. Immediately after heating to the γ - phase, the thermal conductivity increased sharply, similarly to what was observed in [3]. These data were irreproducible. After annealing at 52 K for an hour we obtained curve 4, and it was reproducible both on heating and cooling. The thermal conductivity of nearly free sample 4 decreases slightly as the temperature rises. The isochoric thermal conductivity of the higher-density samples decreases rather sharply in the two-phase region and increases in $\gamma$ - $O_2$ though the increase is much weaker than in [3]. In our measurements the temperature coefficient of the thermal conductivity growth $f=(\partial \ln \Lambda/\partial \ln T)_V$ is about 0.4 which is close to the values obtained for other molecular crystals [7]. The Bridgman coefficient $g = -(\partial \ln \Lambda/\partial \ln V)_T$ calculated from the experimental values is 3.8±0.5 at T=55 K.

As was mentioned previously above 35 K the phonon mean free path estimated by a simple gas-kinetic formula is about $10^{-7}$ cm [3], which is close to the lattice constant. In this case the applicability of the phonon concept becomes questionable. Figure 2 shows the lower limit of the thermal conductivity of the lattice $\Lambda'_{min}$ (broken curve) for sample 3 which was calculated assuming that all the modes are «diffusive» and taking into account the site-to-site transfer of the rotational energy of the molecule [9]:

$$\Lambda'_{min} = (3+z)\left(\frac{\pi}{6}\right)^{1/3} n^{2/3} k_B v \left(\frac{T}{\Theta_D}\right)^2 \int_0^{\Theta_D/T} \frac{x^3 e^x}{(e^x - 1)^2} dx \qquad (1)$$

where $\Theta_D = v(\hbar/k_B)(6\pi^2 n)^{1/3}$, n is the number of atoms (molecules) per unit volume, v is the polarization–averaged sound velocity, z is the number of the rotational degrees of freedom of the molecules. The experimental values of the thermal conductivity are only 30% higher than the lower limit of the thermal conductivity of the lattice $\Lambda'_{min}$. This means that in $\gamma$-$O_2$ most of the heat is transported by «diffusive» modes whose mean free paths are considerably limited.

Low value of the Bridgman coefficient in $\gamma$- $O_2$ agrees with the above assumption. In the general case the Bridgman coefficient $g = -(\partial \ln \Lambda/\partial \ln V)_T$ is the weighted average

over the phonons and «diffusive» modes with significantly different volume dependences [9]:

$$g = \frac{\Lambda_{ph}}{\Lambda} g_{ph} + \frac{\Lambda_{dif}}{\Lambda} g_{dif} , \qquad (2)$$

Since $g_{ph} \sim 3\gamma + 2q - 1/3$, where $q = (\partial \ln\gamma/\partial \ln V)_T \sim 1$ and $g_{dif} \sim \gamma + 1/3$ ($\gamma$ is the Gruneisen constant equal to 2.8 for $\gamma$-$O_2$), Eq. (2) describes the general tendency of the Bridgman coefficient to decrease as increasingly more heat is being transported by the «diffusive» modes.

The concept of the «lower limit» of thermal conductivity can explain the «saturation» of the thermal conductivity but not its growth. Earlier, the increase of the isochoric thermal conductivity with temperature was observed in orientationally disordered phases of some molecular crystals [6-7]. In $\gamma$-$O_2$ this effect can be attributed to a decay of phonon scattering at molecular rotational excitations and short-range magnetic order fluctuations at rising temperature (see [3]).

**Conclusions**

The isochoric thermal conductivity of $\gamma$-$O_2$ has been investigated on samples of varying density in the temperature interval from 44 K to the onset of melting. The thermal conductivity of nearly free sample decreases slightly with rising temperature while it increases at isochoric conditions. The estimates show that absolute value of the thermal conductivity is close to its lower limit and a considerable part of the heat in $\gamma$-$O_2$ is transported by the «diffusive» modes. The growth of the isochoric thermal conductivity may be connected with the decrease of the «rotational» component of thermal resistance due to attenuation of the rotational and magnetic correlations in the motion of the neighboring molecules.

Table 1. Molar volumes $V_m$, temperatures $T_0$ (onset of single-phase region) and $T_m$ (onset of melting).

| Sample № | $V_m$ см³/mole | $T_0$ K | $T_m$ K |
|---|---|---|---|
| 1 | 22.3 | 51 | 83 |
| 2 | 22.45 | 49 | 80 |
| 3 | 22.95 | 45 | 68 |
| 4 | 23.4 | 52 | 56 |

**Figure Captions**

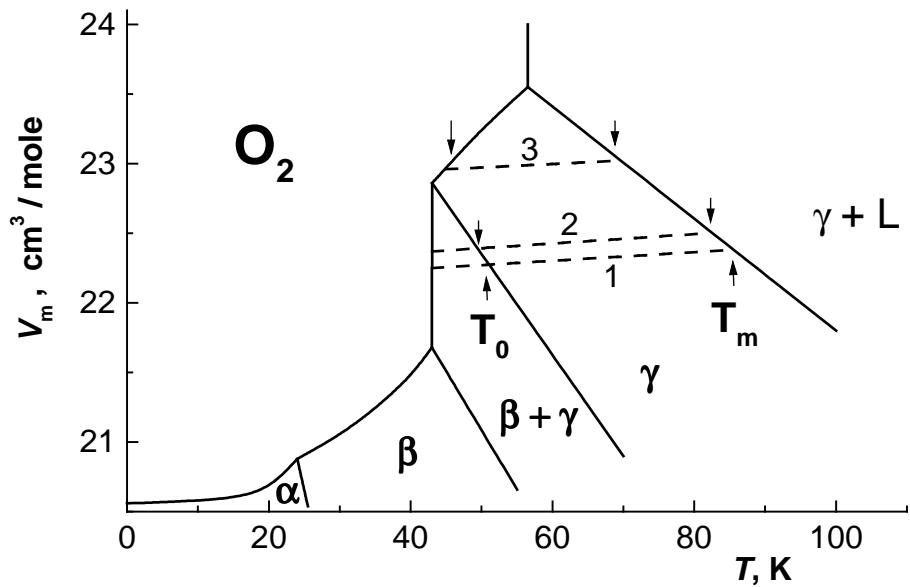

Fig.1. V-T phase diagram of solid $O_2$ plotted using data of [1,2]: (----) - molar volumes; arrows show the onset of single-phase region and melting.

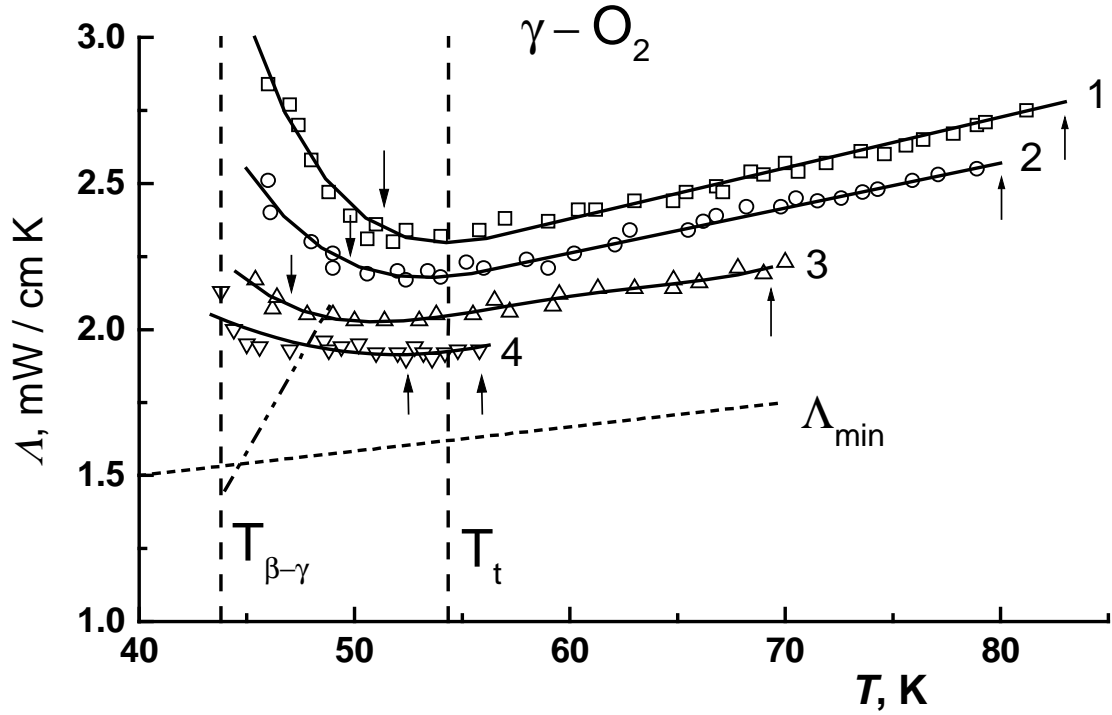

Fig.2. Thermal conductivity of nearly free sample 4 grown under pressure ≈5 MPa ($V_m$=23.4 cm$^3$/mole) and isochoric thermal conductivity of three samples of varying densities (see Table.1) : (——) - smoothed values of isochoric thermal conductivity, (----) - thermal conductivity measured under saturated vapor pressure according to [3]; arrows mark the onset of single-phase regions and the onset of melting; (·····) - lower limit of thermal conductivity of the lattice $\Lambda'_{min}$ for sample 3 calculated assuming «diffusivity» of all the modes and allowing for the site-to-site transfer of the rotational energy of molecules [9].